\documentclass[twocolumn,showpacs,preprintnumbers,amsmath,amssymb]{revtex4}
\usepackage{graphicx}
\usepackage{dcolumn}
\usepackage{bm}
\usepackage{mathrsfs}

\begin{document}

\preprint{APS/123-QED}

\title{Optical vortex soliton with self-defocusing Kerr-type nonlocal nonlinearity}

\author{Shigen Ouyang}
\author{Qi Guo}
\email{guoq@scnu.edu.cn} \affiliation{Laboratory of Photonic
Information Technology, South China Normal University Guangzhou,
510631, P. R. China}

\date{\today}

\begin{abstract}
We develop one numerical method to compute the optical vortex
soliton with self-defocusing Kerr-type nonlocal nonlinearity. With
the numerical simulation method, the propagation and interaction
properties of such optical vortex solitons are investigated.
\end{abstract}

\pacs{42.65.Tg~,~42.65.Jx~,~42.70.Nq~,~42.70.Df}

\maketitle

\section{Introduction}
Optical vortex solitons are self trapped intensity dips with screw
phase dislocations. Owing to such screw phase dislocations, there
is one phase singular point of the vortex whose real and imaginary
parts both vanish. Circumnavigating the phase singular point in a
counterclockwise direction a $\pm 2m\pi$ phase ramp is picked up.
The integer $\pm m$ is called a topological charge of the vortex.
Optical vortex solitons can be generated in self-defocusing
nonlinear medium as consequences of the counterbalanced effect of
diffraction and nonlinearity of the medium. The numerical forms or
approximate forms of the vortices are investigated\cite{14,15,16}.
Optical vortices have been observed in a Kerr type nonlinear
medium\cite{1}, in a photorefractive crystal\cite{2} and in a
saturable nonlinear medium\cite{3,4}. The stability of two
dimensional vortices and one dimensional dark solitons are
investigated\cite{2,4,5,6,7,8,9,13,15}. It is indicated that
one-dimensional dark soliton stripes can decay into optical vortex
solitons due to transverse modulation instability\cite{5,6,9}. A
vortex of charge $|m|>1$ is found to be topologically unstable and
will split into $|m|$ singly charged vortices under perturbations
with the total charge conserved\cite{2,7}. Hight order screw
dislocation ($|m|>1$) are very sensitive to perturbations\cite{7}.
However if the perturbation is very small, these resulted singly
charged vortices will locate too close to be distinguished, and in
this case the splitting multicharged vortex is still viewed as a
whole one formed as these nearly superposed singly charged
vortices. It is indicated that multicharged vortices are very
long-living objects and called to be metastable\cite{8}. The
rotation of a pair of first-order screw dislocation with equal
signs and annihilation of a pair of dislocations of opposite signs
were detected\cite{7,10}. In the limit that the interval distances
between vortices are much larger than the size of the core of the
vortices, the vortices can be viewed as point
vortices\cite{11,12}. Interestingly upon breakup of the input
high-charge vortex the resulting charge-one vortices repel each
other and form an array aligned perpendicular to the anisotropy
axis of photorefractive crystal\cite{2}.

In this paper, We develop one numerical method to compute the
optical vortex soliton with self-defocusing Kerr-type nonlocal
nonlinearity. Such numerical method is similar to those presented
in reference\cite{17,18,19}. With the numerical simulation method,
the propagation and interaction properties of such optical vortex
solitons are investigated.

\section{Numerical method to compute nonlocal vortex soltions}
The propagation of an optical beam in spatially nonlocal
self-defocusing media can be described by this following (1+2)
dimensional nonlocal nonlinear Sch\"{o}dinger equation(NNLSE)
\begin{equation}\label{1-1}
i{{\partial u}\over{\partial z}}+{{1}\over{2}}\nabla_\bot^2u-u\int
R(|\textbf{r}-\textbf{r}^\prime|)|u(\textbf{r}^\prime,z)|^2d^2\textbf{r}^\prime=0
\end{equation}
where $u(\textbf{r},z)$ is the complex amplitude envelop of the
light beam, $|u(\textbf{r},z)|^2$ is the light intensity,
$\textbf{r}=x\hat{\textbf{x}}+y\hat{\textbf{y}}$ and $z$ are
transverse and longitude coordinates respectively,
$\nabla_\bot^2={{\partial^2}\over{\partial
x^2}}+{{\partial^2}\over{\partial y^2}}$ is the transverse Laplace
operator. $R(|\textbf{r}|), (\int R(|\textbf{r}|)d^2\textbf{r}=1)$
is the real axisymmetric nonlocal response function, and $
n(\textbf{r},z)=-\int
R(|\textbf{r}-\textbf{r}^\prime|)|u(\textbf{r}^\prime,z)|^2d^2\textbf{r}^\prime$
is the light-induced perturbed refractive index. Note that not
stated otherwise all integrals in this paper will extend over the
whole transverse x-y plane. When $R(|\textbf{r}|)=\delta(x,y)$,
equation~(\ref{1-1}) will reduce to the local nonlinear
Sch\"{o}dinger equation(NLSE)
\begin{eqnarray}\label{1-2}
i{\partial u\over\partial z}+{{1}\over{2}}\nabla_\bot^2u-|u|^2u=0,
\end{eqnarray}
which has vortex soliton solutions in the form of\cite{3,8,11}
\begin{eqnarray}
u(\textbf{r},z)=F(r)e^{i\beta
z+im\varphi}=\Psi(r,\varphi)e^{i\beta z},\label{1-3}
\end{eqnarray}
where $\beta^*=\beta$ is the phase constant, $\varphi$ is the
azimuthal angle, $m=\pm1,\pm2,\cdots$ is the topology charge,
$F^*(r)=F(r)$, $F(0)=0$ and $F(r)\rightarrow\rm{const}=\eta$ as
$r\rightarrow\infty$.

In this paper we numerically compute the vortex soliton solutions
of NNLSE~(\ref{1-1}) expressed in the form of Eq.~(\ref{1-3}).
Substituting Eq.~(\ref{1-3}) into (\ref{1-1}) and considering the
asymptotic behavior under $r\rightarrow\infty$, we have
$\beta=-\eta^2$. Then we have
\begin{eqnarray}\label{1-4}
\eta^2\Psi+{{1}\over{2}}\nabla_\bot^2\Psi-\Psi\int
R(|\textbf{r}-\textbf{r}^\prime|)|\Psi(\textbf{r}^\prime)|^2d^2\textbf{r}^\prime=0
\end{eqnarray}
We discrete the function $\Psi(x,y)$ in
$\Psi_{jk}=\Psi\big(-h+(j-1)\triangle x,-h+(k-1)\triangle y\big)$,
where $-h<x<h, -h<y<h$ is the sample window, $\triangle
x=\triangle y$ is the sample step. Define the discrete Fourier
transform (DFT) $\mathscr{F}$ by
\begin{subequations}
\begin{eqnarray}
&&\widetilde{\Psi}_{jk}=\mathscr{F}[\Psi]_{jk}=\sum_{p,q=1}^n\rm{F}_{jp}\Psi_{pq}\rm{F}_{qk}\\
&&\Psi_{jk}=\mathscr{F}^{-1}[\widetilde{\Psi}]_{jk}=\sum_{p,q=1}^n\rm{F}^*_{jp}\widetilde{\Psi}_{pq}\rm{F}^*_{qk}
\end{eqnarray}
\end{subequations}
where
$\rm{F}_{jk}={{1}\over{\sqrt{n}}}\exp[i{{2\pi}\over{n}}(j-1)(k-1)]$
and $n={{2h}\over{\triangle x}}+1$. Taking the DFT on
Eq.~(\ref{1-4}),we have
\begin{eqnarray}\label{1-5}
&&\eta^2\widetilde{\Psi}_{jk}-{{1}\over{2}}\Omega_{jk}\widetilde{\Psi}_{jk}\nonumber\\
&&-\mathscr{F}\left[\Psi\int
R(|\textbf{r}-\textbf{r}^\prime|)|\Psi(\textbf{r}^\prime)|^2d^2\textbf{r}^\prime\right]_{jk}=0
\end{eqnarray}
where
$\Omega_{jk}=\left({{2\sin[{{\pi}\over{n}}(j-1)]}\over{\triangle
x}}\right)^2+\left({{2\sin[{{\pi}\over{n}}(k-1)]}\over{\triangle
y}}\right)^2$. From Eq.~(\ref{1-5}), we obtain
\begin{eqnarray}\label{1-6}
\widetilde{\Psi}_{jk}&&={{(\mu+\eta^2)\widetilde{\Psi}_{jk}-\mathscr{F}\left[\Psi\int
R(|\textbf{r}-\textbf{r}^\prime|)|\Psi(\textbf{r}^\prime)|^2d^2\textbf{r}^\prime\right]_{jk}}\over{\mu+{{1}\over{2}}\Omega_{jk}}}\nonumber\\
&&\equiv\mathscr{D}[\Psi]_{jk}
\end{eqnarray}
where $\mu$ is an arbitrary positive constant.

We use Eq.~(\ref{1-6}) to iteratively compute the vortex soliton
solutions. For an initial trying function, for instance,
\begin{eqnarray}
\Psi_0(r,\varphi)=\eta\left[1-\exp\left(-{{r^m}\over{\sigma^m}}\right)\right]\exp(im\varphi),
\end{eqnarray}
where $\sigma$ is a constant, from Eq.~(\ref{1-6}), we get
$\widetilde{\Psi}_1=\mathscr{D}[\Psi_0]$ and
$\Psi_1=\mathscr{F}^{-1}[\widetilde{\Psi}_1]$. For $p\geq1$, we
get the iteration scheme
$\widetilde{\Psi}_{p+1}=\mathscr{D}[\Psi_p]$ and
$\Psi_{p+1}=\mathscr{F}^{-1}[\widetilde{\Psi}_{p+1}]$. Perform the
iteration until some accuracy is achieved, then we get the
approximate numerical vortex soliton solutions.
\begin{figure}
\centering
\includegraphics[totalheight=1.3in]{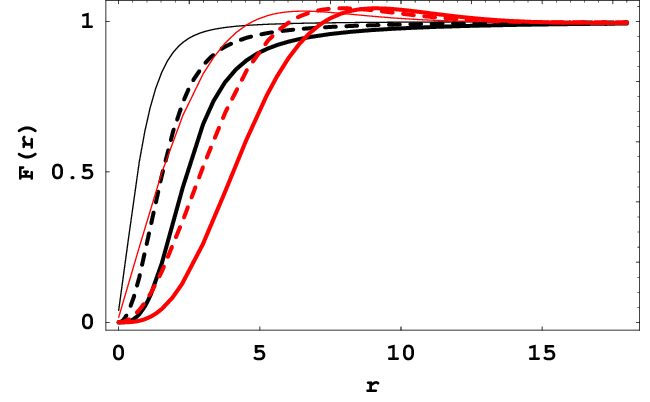}
\caption{\label{fig01}Modulus of vortex' amplitude $F(r)$ with
$F(\infty)=\eta=1$:thin solid dark line corresponds to local case
$w=0$ and charge $m=1$; thin solid red line $w=6,m=1$; dashing
dark line $w=0,m=2$; dashing red line $w=6,m=2$; thick solid dark
line $w=0,m=3$; thick solid red line $w=6,m=3$.}
\end{figure}
\begin{figure}
\centering
\includegraphics[totalheight=1.7in]{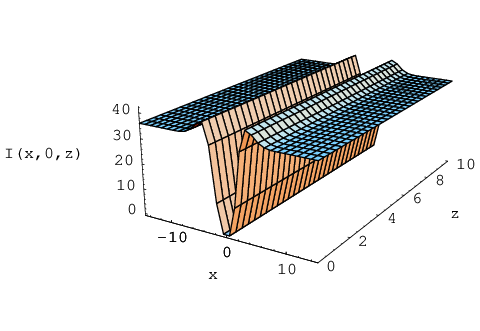}
\caption{\label{fig02}Intensity $I(x,0,z)$ on plane $y=0$ of
vortex of charge $m=2$ obtained by numerical simulation with
parameters $\eta=6,w=10$.}
\end{figure}

As an example, we consider this following nonlocal case in which
the light-induced perturbed refractive index is governed by
\begin{eqnarray}
n-w^2\nabla^2_\bot n=-|u|^2,
\end{eqnarray}
which results in
\begin{equation}
n(\vec{r},z)={{1}\over{2\pi w^2}}\int
\rm{K}_0\left({{|\vec{r}-\vec{\rho}|}\over{w}}\right)|u(\vec{\rho},z)|^2d^2\vec{\rho},
\end{equation}
where $\rm{K}_0(x)$ is the modified Bessel function of the second
kind and $w$ is the characteristic nonlocal length. As indicated
in Fig.~(\ref{fig01}), the cores of vortices of the same topology
charge and the same background intensity increase with the
increasing of characteristic nonlocal length. The numerical
simulation shown in Fig.~(\ref{fig02}) indicates that the
numerical solution of vortices obtained by our method can describe
the vortices very well. As shown in Fig.~(\ref{fig02}) there is no
observable splitting of the vortex of charge $m=2$ with absence of
perturbations during the numerical simulation length $z=10$. It is
consistent with the statement that multicharged vortices are
metastable\cite{8}.

\begin{figure}
\centering
\includegraphics[totalheight=1.5in]{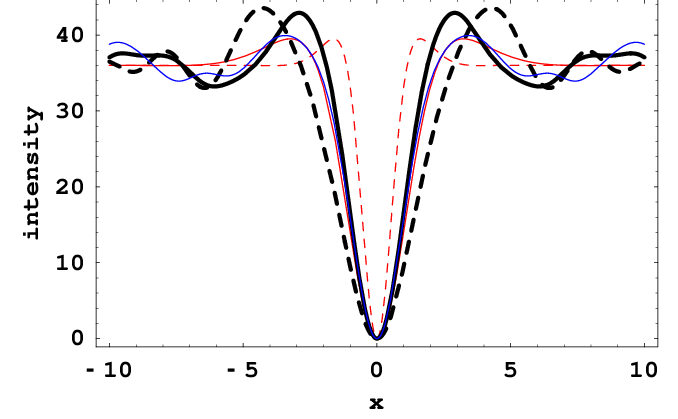}
\caption{\label{fig08}Intensity profiles on plane $y=0$ of the
evolving singly charged vortex with an initially shrunk core size.
Dashing red line corresponds to input intensity of input beam with
half of the core size of the corresponding vortex soliton; Dashing
dark line corresponds to intensity at $z=3.3$; Solid dark line
$z=8.7$; Blue line $z=17.5$. The solid red line is the intensity
of the corresponding singly charged vortex soliton. The simulation
parameters are $m=1, \eta=6, w=10$.}
\end{figure}

\begin{figure}
\centering
\includegraphics[totalheight=1.3in]{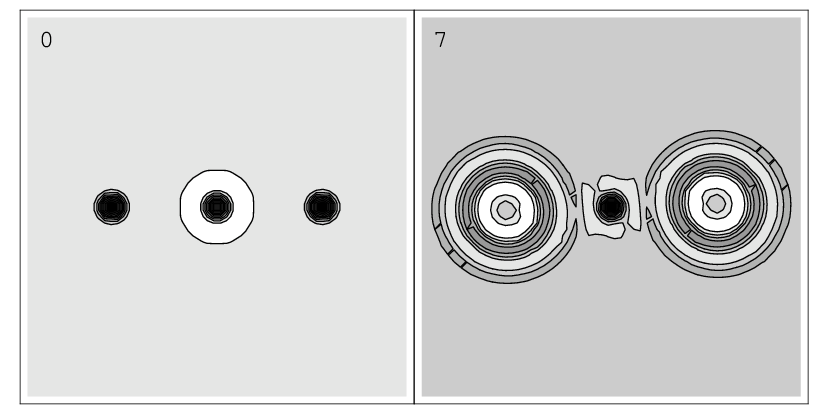}
\caption{\label{fig03}Numerical simulation under initial condition
described by Eq.~(\ref{3-01}). The left figure shows input
intensity at $z=0$, a singly charged vortex located at center and
two side-locating normal Gaussian dips initially separated by a
distance $d=30$. The right figure shows intensity at $z=7$. The
simulation parameters are $w=10,\eta=6, d=30$ and the size of each
frame is equal to $27\times27$ unit.}
\end{figure}

To numerically investigate the stability of singly charged vortex,
we use an input vortex with half of the core size of the
corresponding vortex soliton, that is
\begin{eqnarray}
u(r,\varphi,0)=F(2r)\exp(i\varphi),
\end{eqnarray}
where $F(r)$ is the modulus of the corresponding vortex soliton's
amplitude. The simulation results are shown in Fig.~(\ref{fig08}),
from which we can find the initially shrunk singly charged vortex
will evolve to the corresponding vortex soliton along with
radiation of ripples. So it is numerically implies that singly
charged vortex is stable.
In another case we consider this following initial amplitude
\begin{eqnarray}\label{3-01}
u(x,y,0)&&=\left[1-e^{-{{(x-d/2)^2+y^2}\over{2}}}-e^{-{{(x+d/2)^2+y^2}\over{2}}}\right]\nonumber\\
&&\times F(x,y)e^{i\varphi(x,y)},
\end{eqnarray}
which describe a field consisting of a singly charged vortex
located at $(x=0,y=0)$ and two normal Gaussian dips located at
$(d/2,0)$ and $(-d/2,0)$ respectively. The numerical solution of
Eq.~(\ref{1-1}) under initial condition (\ref{3-01}) is shown in
Fig.~(\ref{fig03}), from which, we can find that the two initial
normal Gaussian dips without screw phase dislocation seeded will
radiate ripples and become wider and wider, whereas, the singly
charged vortex maintains its shape during propagation.

As has been previously indicated\cite{7,10} that two separated
vortices of the same charge $m=1$ embedded off axis in a
finite-size Gaussian beam will rotate around the axis in both the
linear and nonlinear cases. The rotation angular speed depends on
the propagation distance from the beam waist of the background
Gaussian beam and is not a constant. There exits a maximal
rotational angle at infinite propagation distance from the beam
waist. In this paper we investigate the rotation of vortices
embedded in an infinite-size constant background $\eta$. To do so,
we consider this following initial amplitude
\begin{eqnarray}\label{3-02}
u(x,y,0)&&={{1}\over{\eta}}F_1(x-d/2,y)e^{i\rm{m}_1\varphi(x-d/2,y)}\nonumber\\
&&\times F_2(x+d/2,y)e^{i\rm{m}_2\varphi(x+d/2,y)},
\end{eqnarray}
which describes two vortices of charge $m_1$ and $m_2$ initially
located at $(d/2,0)$ and $(-d/2,0)$ respectively.
\begin{figure}
\centering
\includegraphics[totalheight=1.1in]{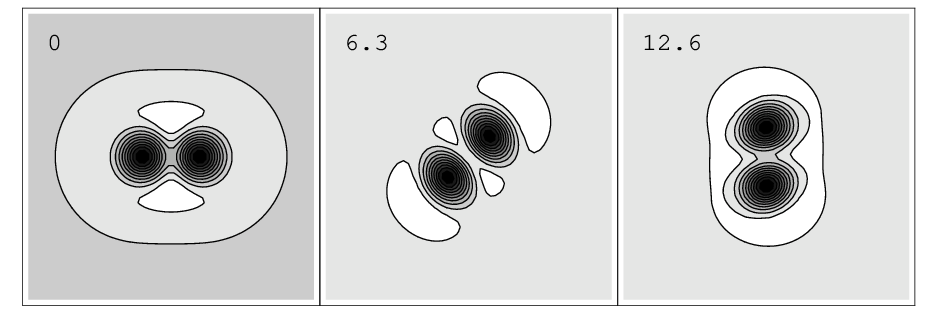}
\caption{\label{fig04}Rotation of two vortices of the same charge
$m=1$ initially separated by a distance $d=4$, and $w=10,\eta=6$.
The size of each frame is equal to $10\times10$ unit.}
\end{figure}
As shown in Figs.~(\ref{fig04}) the two vortices of the same
charge $m=1$ rotate around each other. The rotational angular
speed is nearly a constant and there is no limit on the rotational
angle of the two vortices. As indicated by table(\ref{table01}),
the angular speed does depend on the initial separated distance
between vortices.
\begin{table}
\centering \caption{The rotation period $T$ of two vortices of the
same charge $m=1$ initially separated by a distance
$d$.}{\label{table01}}
\begin{tabular}{c|cccccc}
\hline\hline
d & 3 & 4 & 5 & 6 & 7 & 8 \\
\hline
T & 36.1 & 50.4 & 73.9 & 108.8 & 152.4 & 200.8 \\
\hline
$d^2\pi/T$ & 0.78 & 0.997 & 1.06 & 1.04 & 1.01 & 1.001 \\
\hline\hline
\end{tabular}
\end{table}
As has been previously indicated\cite{11,12}, in the limit that
the interval distances between vortices are much larger than the
size of the core of the vortices, the vortices can be viewed as
point vortices. For two vortices of the same charge $m=1$
separated by a very large distance $d$, based on the point
vortices model it can be deduced that the peripheral speed of each
vortex is equal to $1/d$. On the other hand the peripheral speed
can be expressed as ${{d}\over{2}}{{2\pi}\over{T}}$, where $T$ is
the rotation period. So we have ${{d^2\pi}\over{T}}=1$. As
indicated by table(\ref{table01}), the point vortices model can
give a rather good approximation of the peripheral speed of the
vortices for large interval distance.

\begin{figure}
\centering
\includegraphics[totalheight=1.1in]{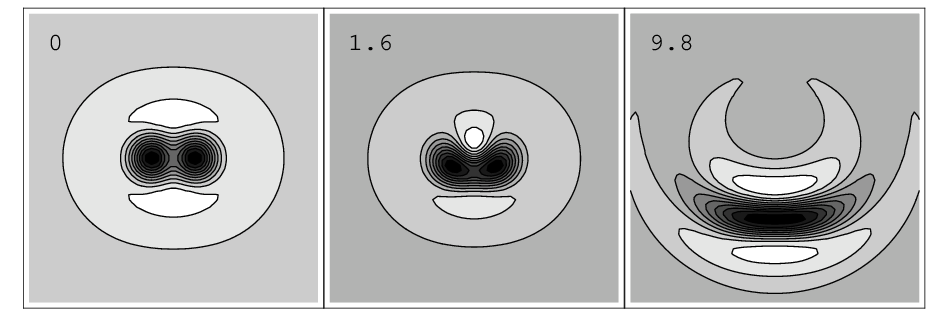}
\caption{\label{fig05}Annihilation of two vortices of opposite
charge $m=1$ and $m=-1$ initially separated by a distance $d=3$,
and $w=10,\eta=6$. The size of each frame is equal to $10\times10$
unit.}
\end{figure}

\begin{figure}
\centering
\includegraphics[totalheight=1.1in]{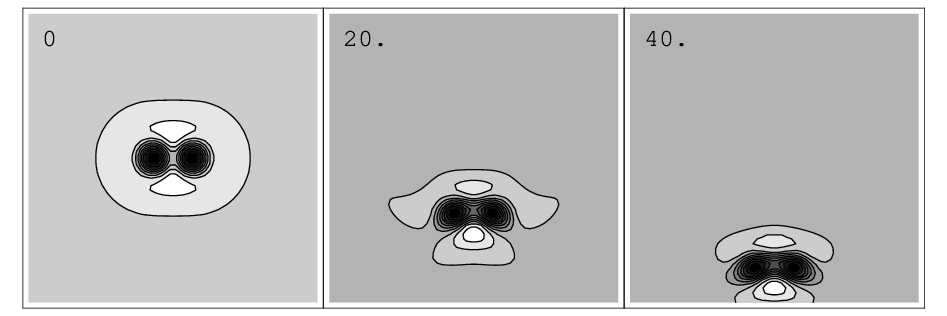}
\caption{\label{fig05a}Co-moving but not annihilated two opposite
vortices initially separated by a distance $d=4$. The other
parameters are the same as Fig.~(\ref{fig05}) and the size of each
frame is equal to $15\times15$ unit.}
\end{figure}

Annihilation of two vortices of opposite charge $m=1$ and $m=-1$
initially separated by a short enough distance are shown in
Fig.~(\ref{fig05}). It is shown that these two opposite vortices
annihilates each other and form a moving trough of the radiated
ripples, which qualitatively agrees with the experimental
observation by I. V. Basistiy, et. al.\cite{7}. It is worth to
note that annihilation for two far separated opposite vortices may
be hardly possible with absence of other exterior interaction.
Figure.~(\ref{fig05a}) may be regarded as a rudimentary proof of
this statement. For large interval distance, the point vortices
model\cite{11,12} predicts the directions of velocities of two
vortices of opposite charge $m=-1$ and $m=1$ are the same and
perpendicular to the line along these two vortices. So these two
far separated vortices will never move close to each other and
annihilation cannot occur. From Fig.~(\ref{fig05a}) the co-moving
speed of vortices is equal to $11.4/40=0.285$, which is very close
to $1/d=1/4=0.25$ predicted by point vortices model.

\begin{figure}
\centering
\includegraphics[totalheight=1.1in]{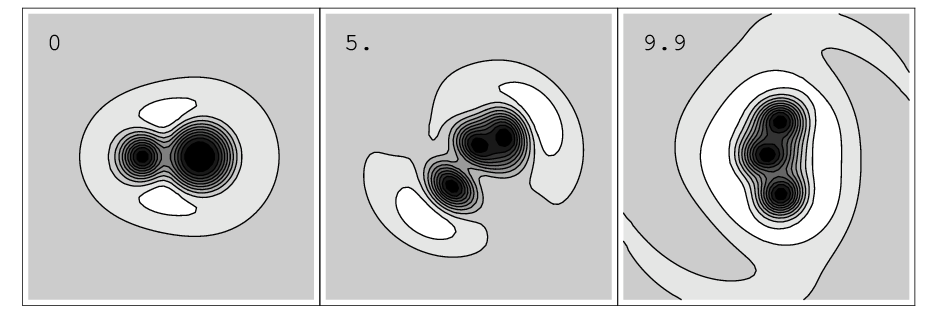}
\caption{\label{fig06}Interaction of two vortices of charge $m=1$
and $m=2$ initially separated by a distance of $d=4$. The other
parameters are the same as Fig.~(\ref{fig04}).}
\end{figure}

\begin{figure}
\centering
\includegraphics[totalheight=1.1in]{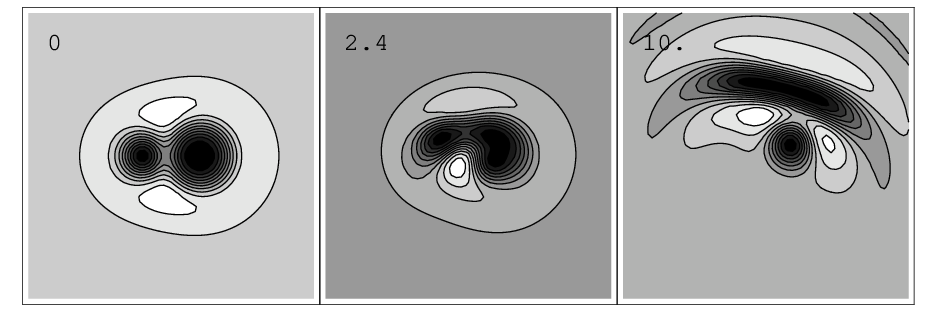}
\caption{\label{fig07}Interaction of two vortices of charge $m=1$
and $m=-2$. The other parameters are the same as
Fig.~(\ref{fig06}).}
\end{figure}

Multicharged vortices are topologically unstable\cite{2,7}. In
reference\cite{2} two possible mechanisms that split a
multicharged vortex into a set of singly charged vortices are
discussed. The first mechanism is due to the fact that
multicharged vortices are topologically unstable and separate into
a set of singly charged vortices in the presence of a small amount
of noise, even in the framework of linear optics\cite{2}. The
second mechanism is due to propagation effects. Anisotropic
initial conditions can result into the splitting of multicharged
vortices due to linear diffraction\cite{2}. Specifically the
reference\cite{2} investigated how an elliptically shaped
high-charge vortex embedded in a Gaussian beam splits into
charge-one vortices. In this paper we consider another anisotropic
initial conditons that a singly charged vortex and a charge-two
vortex initially separated by a distance co-propagate in a
nonlocal media. As shown in Figs.~(\ref{fig06}) and (\ref{fig07}),
the charge-two vortex splits into two singly charged vortices with
the present of another singly charged vortex. We note the
nonlocality of the nonlinear response enhance the anisotropy of
the initial conditions. In local nonlinear case, a local region's
perturbed refractive index is solely generated by the vortex
located at such a region. So in local nonlinear case two far
separated vortices will never affect the perturbed refractive
index of the region the other one located at. So the perturbed
refractive index of the region the charge-two vortex located at
will be isotropic but not anisotropic. The charge-two vortex will
not experience the anisotropy due to the present of another far
separated vortex in the local nonlinear case and will not split
into two singly charged vortices. However in nonlocal nonlinear
case, the perturbed refractive index of a region will depend on a
distant field. The perturbed refractive index of the region the
charge-two vortex located at will be anisotropic due to the
present of another far separated vortex. Owing to such anisotropy
of the perturbed index, the charge-two vortex splits into two
singly charged vortices.

By the way, in \emph{self-focusing} nonlocal case, there exist
vortex soliton solutions which have a vanishing intensities rather
than non-vanishing intensities in self-defocusing case when
$r\rightarrow\infty$. Multicharged vortices are topologically
unstable too in such self-focusing nonlocal case. And we will
predict that the multicharged vortices will also split into singly
charged vortics with the present of far separated vortices.


　　　　　　　　　　　　　　
\end{document}